\documentclass[journal]{vgtc}                     

\onlineid{1531}

\vgtccategory{Research}

\vgtcpapertype{Theoretical \& Empirical}

\title{Average Estimates in Line Graphs Are Biased Toward\\Areas of Higher Variability}

\author{\authororcid{Dominik Moritz}{0000-0002-3110-1053}, \authororcid{Lace M. Padilla}{0000-0001-9251-5279}, \authororcid{Francis Nguyen}{0000-0003-3146-6148}, and \authororcid{Steven L. Franconeri}{0000-0001-5244-9764}}
\authorfooter{
\item
Dominik Moritz is with Carnegie Mellon University. E-mail: domoritz@cmu.edu.
\item
Lace M. Padilla is with Northeastern University. E-mail: l.padilla@northeastern.edu.
\item
Steven L. Franconeri is with Northwestern University. E-Mail: franconeri@northwestern.edu.
\item
Francis Nguyen is with UBC. E-mail: frnguyen@cs.ubc.ca.
}

\abstract{
We investigate \textit{variability overweighting}, a previously undocumented bias in line graphs, where estimates of average value are biased toward areas of higher variability in that line. 
We found this effect across two preregistered experiments with 140 and 420 participants. These experiments also show that the bias is reduced when using a dot encoding of the same series. We can model the bias with the average of the data series and the average of the points drawn along the line. 
This bias might arise because higher variability leads to stronger weighting in the average calculation, either due to the longer line segments (even though those segments contain the same number of data values) or line segments with higher variability being otherwise more visually salient. Understanding and predicting this bias is important for visualization design guidelines, recommendation systems, and tool builders, as the bias can adversely affect estimates of averages and trends. 
} 


\teaser{
  \centering
  \includegraphics[width=\linewidth]{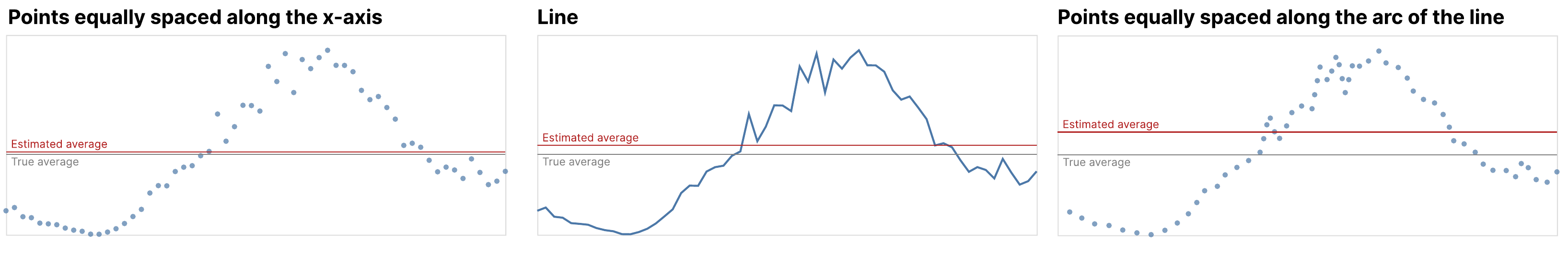}
  \caption{Demonstration of the bias toward variability for three mark types showing the same data. The red line shows the mean estimated averages across all participants in our second experiment. The line chart (center) shows a bias of the estimated average toward higher variability in the higher y-values. The bias is smallest when the data is shown as points equally spaced along the x-axis (left). The bias in line charts is in the same direction as the bias of estimates of points sampled at equal intervals along the arc of the line (right).}
  \label{fig:teaser}
}

\graphicspath{{figs/}{figures/}{pictures/}{images/}{./}} 

\usepackage{tabu}                      
\usepackage{booktabs}                  
\usepackage{lipsum}                    
\usepackage{mwe}                       

\usepackage{mathptmx}                  

\usepackage{scrextend} 
\usepackage{tabularray} 

\usepackage[normalem]{ulem}

\newcommand{\added}[1]{#1}
\newcommand{\removed}[1]{}

\usepackage{amsmath}

\usepackage{xspace,xpunctuate}
\newcommand{\ie}{{i.e.,}\xspace}
\newcommand{\eg}{{e.g.,}\xspace}
\newcommand{\ea}{{et~al\xperiod}\xspace}

\begin{document}


\firstsection{Introduction}

\maketitle

Since William Playfair invented line graphs in 1786~\cite{playfair}, they have become one of the most common data visualization types. Designers use line graphs to visualize stocks, sensor data, machine learning metrics, and human vitals (\eg heart rate). Line graphs show a continuous variable's change over another continuous variable, typically time, as the changing position of a line mark.

We generally assume that visualizations, especially of effective visual encoding channels such as position, are perceived not perfectly but without bias~\cite{cleveland}. The popularity of line graphs may be because the visual encoding of time series as the position of a line is considered effective relative to other visual encoding channels, such as hue, depending on the task. 
However, designers should be cognizant of perceptual biases that can lead to misinterpretation of visualizations~\cite{huff2023lie, tufte1985visual}. For example, prior work demonstrates that the background color can bias the perception of the color of marks~\cite{szafir2018good}, and continuous rainbow color maps are perceived as discrete categories~\cite{liu2018somewhere, quinan2019examining}. 

There may be unexplored biases in line charts as well.
 When drawing a line, the length of the line drawn varies not only with the duration of the visualized time series but also with the variability of the values (and the resulting variability of the line graph). For example, take two time series of regularly sampled values over the same duration. The first value may be constant while the second value oscillates. Both time series have the same number of values (the same duration), but in the visualization as a line graph, the second line has a longer overall length---we call this the \emph{arc length} of the line. The arc length is the sum of the length of all line segments. Steeper line segments are longer than other line segments of the same length along x. The arc length of a line affects how much visual weight a line has (how much ``ink'' is needed to draw it) and \added{how much it draws viewers' attention~\cite{salience}}. Within a single line, periods of the same length may have a longer or shorter arc length depending on how much the line goes up and down, which depends on the amount of variability in the visualized time series.

Estimates of average values may be biased by design features of the marks that draw viewers' attention, as found in prior work~\cite{hong2021weighted}, and increased variability in visualized times series may capture attention.
~\added{Our bottom-up attention is generally attracted to visual information that contrasts with its surroundings~\cite{salience}. Marks can vary in contrast to the background and other elements, which dictates how capturing they are to our attention, referred to as \textit{salience}. For example, areas of a line graph with high variability also have more ink (often in color) and more edges, creating high contrast with the background. }
Therefore, we hypothesize that average estimates in lines are biased toward areas of line graphs that have a longer relative arc length (\ie that have a longer arc length for the same duration or that use more ink). Put differently, we hypothesize that increased variability in higher values increases the average estimate of a time series (and vice versa) in line graphs and that the bias is consistent with the salience of the line.

We tested this hypothesis in two experiments. Our first experiment showed that average estimates are biased toward the area of the line that visualizes more variable data. In the second experiment, we sought to understand the reasons for the observed bias. We hypothesized that average estimates in line graphs are consistent with the salience of a line. We, therefore, hypothesized that average estimates of points drawn along the arc of a line are more consistent with average estimates on lines than points drawn at regular intervals along the time dimension of a graph. In other words, the bias to variability may decrease from a line graph encoding to a point encoding of the same data. The results of the second experiment confirm this hypothesis, and a demonstration of the findings is shown in~\Cref{fig:teaser}. We preregistered the experiments on the Open Science Framework (\href{https://osf.io/j2vn3}{Experiment 1} and~\href{https://osf.io/7h2zy}{Experiment 2}
) and have made our study materials available at (\href{https://osf.io/aupbk/}{OSF Link}).

Although we are not the first to document that different time series have different arc lengths, and that arc length affects aggregates~\cite{million,cpc}, we experimentally show the effect and reveal how much average estimates in line graphs can be biased by variability in the time series. Understanding this bias is important because people often use line graphs (as one of the most common visualization types) to visually assess whether values are, on average, above or below critical thresholds or to estimate future trends. Our results show that variability in \removed{the} line graph biases the perception of averages and, therefore, conclusions people draw. While the variability affects audiences' perception, it may be an artifact of an irrelevant factor that should not affect their conclusions. We discuss these implications and potential designs that could reduce the observed bias.


\section{Related Work}

Line graphs are a common visualization--- especially of time series data---in various domains~\cite{quadri2021survey}, appearing in papers, reports, monitoring dashboards, and visual analysis systems. They are generally considered an effective visualization for time series data~\cite{waldner2019comparison}. Mackinlay describes a visualization as \emph{effective} when the information it conveys is more readily perceived than with other visualizations~\cite{apt}. A visualization is always effective only with respect to a particular \emph{task}. In this paper, the task is to estimate the average of a time series, which corresponds to ``compute derived value'' in Amar \ea's popular low-level task taxonomy~\cite{amar}. Whether a visualization is effective depends on choosing the right visual mark and effective visual encoding channels. Position is considered the most precise visual encoding channel~\cite{cleveland}.

However, research also demonstrates the limitations of line graphs for various tasks~(\eg \cite{gogolou2018comparing, albers2014task, correll2012comparing, biased_average, cleveland, adnan2016investigating, heer2009sizing, javed2010graphical}; reviewed in~\cite{quadri2021survey}). Several studies compared line charts to other encodings and found that the efficacy of line charts depends on the task~\cite{albers2014task, correll2012comparing,heer2009sizing, javed2010graphical}. 
Albers, Correll, and Gleicher found that line graphs are best suited for identifying the min, max, and range while less effective for average estimation~\cite{albers2014task} (see also~\cite{correll2012comparing}). 
Studies have shown that positional encoding may be a precise visual encoding channel, but it can produce systematic biases regarding how averages are perceived~\cite{biased_average} and remembered~\cite{mccoleman2021rethinking}. Researchers investigated bias in composed displays with line and bar graphs~\cite{biased_average}. 
When comparing two curved lines, work shows that the steepness of the lines causes a perceptual illusion making it challenging to estimate differences between the lines visually~\cite{cleveland}. 

We are not the first to recognize that line graphs dedicate 
\added{more visual weight} to steeper lines. This effect becomes critical when summarizing large ensembles of line graphs. Heinrich and Weiskopf reduced the salience of steeper lines in density visualizations of parallel coordinate plots~\cite{cpc}. Moritz and Fisher aggregated line graphs to create density visualizations of large time series~\cite{million}. To avoid visual artifacts of steeper lines, they normalized each line by the arc length such that each time series contributes equally to the density visualization. Zhao \ea proposed an effective density computation and extended density visualizations with interactivity~\cite{zhao2021kd}. However, none of these works experimentally confirm that average position estimates in line graphs are biased due to the increased salience.


\section{Experiments}
In two experiments, we investigated the perception of average values in line graphs. The goal of the first experiment was to determine if the perception of averages is biased toward variability and whether we can predict this bias. 
To examine one possible source of the bias, in Experiment 2, we aimed to identify the contribution of the line encoding. To test this, we conducted a study comparing the bias of three mark types: 1) points equally spaced along the x-axis, or \emph{Cartesian spaced}, 2) points equally spaced along the arc of the line, and 3) a line.


\subsection{Experiment 1}
The first experiment aimed to determine if there was a previously
undocumented bias in perceptual line graph average estimation. We termed this potential bias \textit{variability-overweighting}, which is when people believe that the average value of the data set is closer to the more variable data. For example, in \Cref{fig:teaser} \added{center}, if a participant were to indicate that the average value was located at the red line, they would be incorrect. In the figure, the \added{true} average is lower, but it could be the case that people are biased toward the more variable data. 

We hypothesized that individuals would have a skewed perception of averages toward sets of values with higher variance. High variance increases the amount of ink in a line graph and, therefore, the visual saliency of the line. If, in a line graph, the variance correlates with the value plotted along the same axis (here y), then we expect people to estimate that most values are where the high-variance data is located. 

\begin{figure}[!t]
  \centering
  \includegraphics[width=\columnwidth]{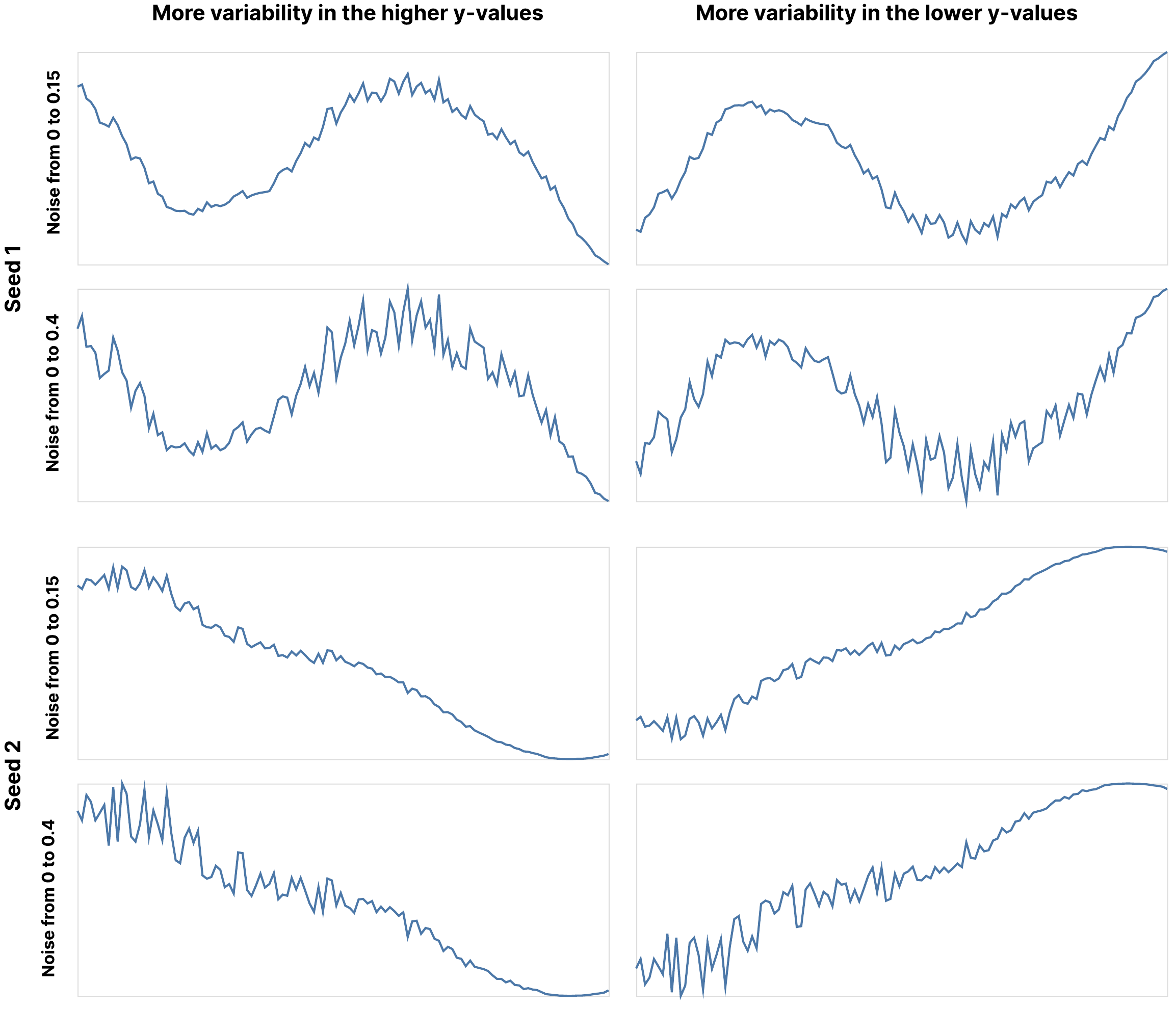}
 \caption{First 8 of 48 stimuli for Experiment 1. The stimuli included two variability levels for each seed and conditions where the graphs were mirrored creating stimuli with variability in the higher and lower y-values.}
  \vspace{-.5em}
  \label{fig:exp_1_stimuli}
\end{figure}

\begin{figure}[!t]
  \centering
  \includegraphics[width=0.85\columnwidth]{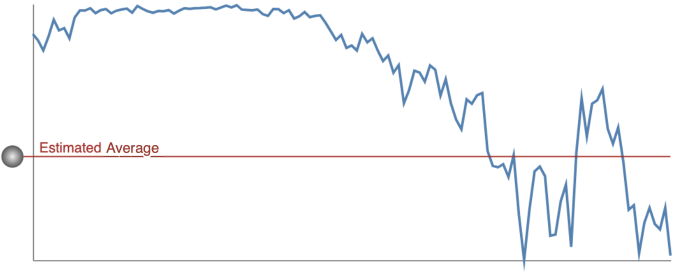}
 \caption{Example stimulus. Participants can move the grabber up and down to where they estimate the line's average to be. Each series is shown as a 500px by 200px graph.}
  \vspace{-.5em}
  \label{fig:prereg_exp}
\end{figure}

To test if variability-overweighting occurs with line graphs, in Experiment 1, we showed participants line graphs of synthetic stock data that we modified to induce increased variability. We created stimuli that included more variability in the higher y-values and then reflected the stimuli to create graphs with more variability in the lower y-values (see~\Cref{fig:exp_1_stimuli}). We will refer to the stimuli with variability in the higher y-values as ``variability upper'' and those with variability in the lower y-values as ``variability lower.'' Participants were tasked with estimating the average y-value of the stock data using a draggable line \added{(\Cref{fig:prereg_exp})}. We used a 2 (variability upper vs. lower) $\times$ 2  (more vs. less variability) within-subjects design for a total of 4 stimuli types of interest. 

We generated the stimuli from 12 seeds to create 48 trials to ensure test-retest reliability. Creating images reflected vertically allowed us to test if variability-overweighting occurs similarly for higher or lower areas on the y-axis. Participants were shown 48 images in a randomized order and estimated the average y-value for each image. We calculated each judgment's Euclidean distance estimation error by subtracting the actual average from the estimated average. The direction of the error was preserved, such that positive values indicated an overestimation of the average value, and negative values represented an underestimation.

\subsection{Experiment 2} 
\begin{figure}[!t]
  \centering
  \includegraphics[width=\columnwidth]{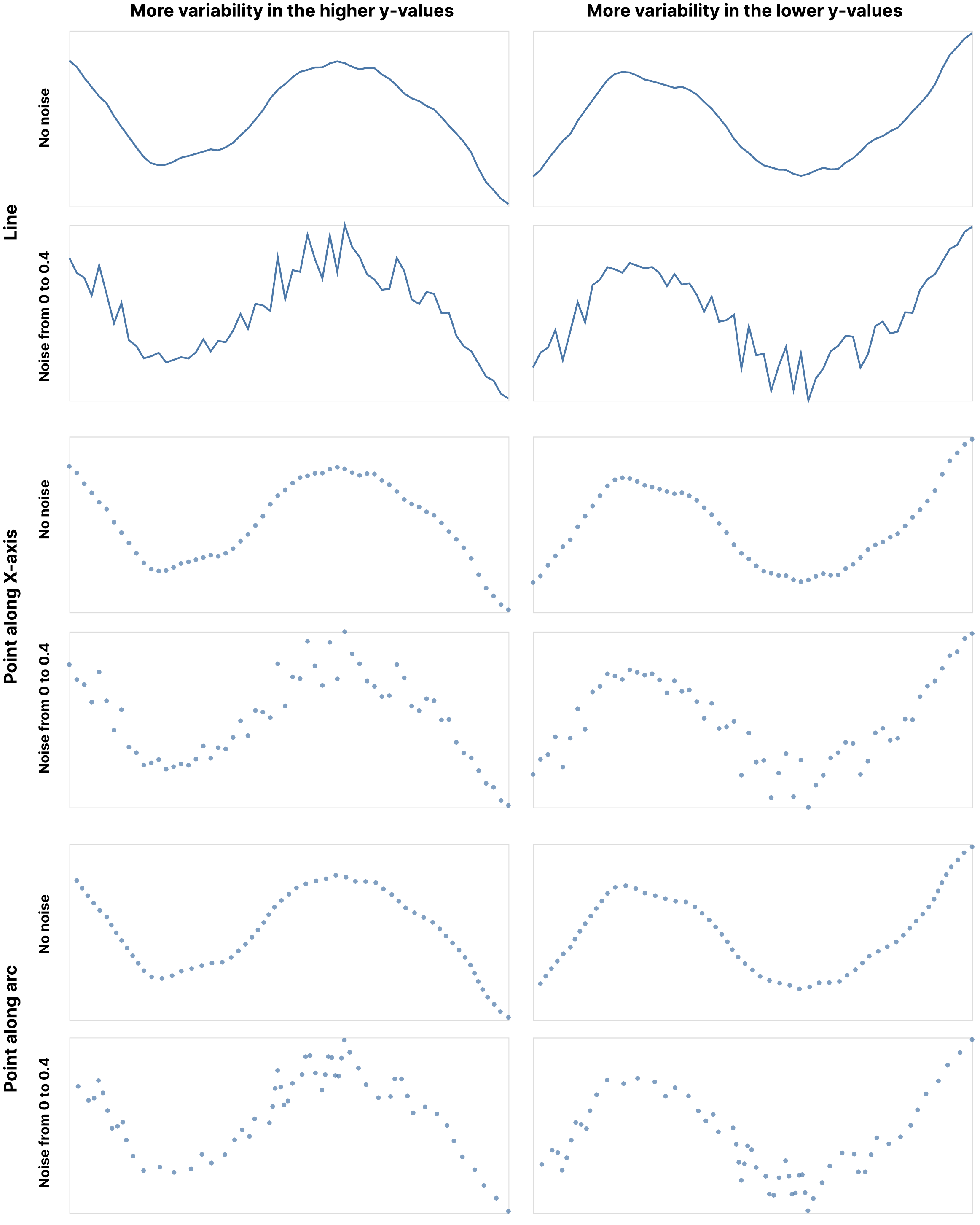}
 \caption{First 8 of 144 stimuli for Experiment 2. For each seed, we included two variability levels, three mark types, and stimuli with variability in the higher and lower y-values.}
  \label{fig:exp_2_stimuli}
\end{figure}

In Experiment 2, we aimed to determine if we could influence the degree of variability-overweighting by changing the mark type. To test this, we replicated Experiment 1, but we encoded the data using 1)~Cartesian spaced points, 2)~points equally spaced along the arc of the line, and 3)~the same line encoding used in Experiment 1 (see \Cref{fig:exp_2_stimuli}).

We hypothesize that by encoding the time series data as \added{Cartesian spaced} points, we can reduce the bias toward more variable data. Each data point is rendered as one point mark without a connection in this encoding. Therefore, two rendered points have the same salience regardless of their distance in y. In lines, the salience of the mark depends on the length of the arc. We also rendered points at equally spaced intervals along the arc (points along the arc) to simulate this behavior in our experiment. With points along arc, the average y-position of the points is heavily biased toward more variability since lines between neighboring points with more different values are longer. Therefore more points are along the arc of lines with more variability. In this encoding, a perfectly accurate viewer cannot estimate the true average of the underlying data series. We also included a line encoding to replicate Experiment 1.

As in Experiment 1, we showed participants graphs of synthetic time series. We then ask them to estimate the average using a draggable line. We used a 3 (point along x, point along arc, line) $\times$ 2 (variability upper vs. lower) $\times$ 2 (more variability vs. no variability) design.

We generated the stimuli types from 12 seeds to create 144 total trials. Participants viewed 48 images of one mark type from the 144 trials, and we calculated the error similarly to Experiment 1. We switched to a between-subject experiment to limit the number of graphs each participant saw and reduce the possibility of potential bias of viewing multiple graph types.

\section{Stimuli generation}

We generated the stimuli for both experiments with the same process. This process used a simulation to generate realistically-looking line charts that we add linearly-interpolated noise to. We re-scaled the series to correct for a subtle yet important bias introduced by the noise.
The code for our stimuli generation for \href{https://observablehq.com/@cmudig/bias-experiment-1}{Experiment 1} and \href{https://observablehq.com/@cmudig/bias-experiment-2}{Experiment 2} are available online and as supplemental material.

\subsection{Experiment 1}

The stimuli (with a sample shown in \Cref{fig:exp_1_stimuli}) are line graphs of randomly generated series of \added{numbers}. Each series has 120 data points. A series is generated from a base series to which we add noise. We generated base series from a geometric Brownian motion stochastic process~\cite{enwiki:1140966364}, a process used to generate realistic-looking stock data. We set $\mu=0$ and $\sigma=1$. We then applied a moving average over $30$ points to smooth the base series. To get 120 data points in a series, we generate $120+30=150$ data points from geometric Brownian motion. We scale the base series, so all values are between zero and one. To the base series (consisting of data points {\color{Teal} $\text{base}_i$}), we added uniform random noise (centered around 0). The amount of noise increases linearly with the value of the base series for positive y-alignment (linear interpolation \added{between {\color{SeaGreen}lowVariability} and {\color{Indigo}highVariability}}).

\vspace{-1em}
\begin{equation}
\begin{split}
    {\color{Maroon}\text{noise}_i} &= \text{\color{SeaGreen} lowVariability} \times (1-{\color{Teal} \text{base}_i}) + \text{\color{Indigo} highVariability} \times {\color{Teal} \text{base}_i} \\
    {\color{FireBrick}\text{dataPoint}_i} &= {\color{Teal} \text{base}_i} + (\text{rand()} - 0.5) \times {\color{Maroon}\text{noise}_i}
\end{split}
\end{equation}
\vspace{-1em}

Low variability series have less variability (0.15) than high variability series (0.4). We seeded the random number generator for the geometric Brownian motion stochastic process and noise to reproduce the same series. We curated the set of seeds to generate diverse line graphs with different shapes.

We generated a series for negative y-alignment by mirroring the series vertically. We mirror each series to understand whether the average estimates may be biased toward higher and lower y-values and counter-balance this bias in our experiments to measure bias toward higher variability.

\vspace{-1em}
\begin{equation}
    \text{mirroredDataPoint}_i = 1- {\color{FireBrick}\text{dataPoint}_i}
\end{equation}

\subsection{Scaling the averages}
\label{sec:scaled-stimuli}

To allow for comparison across stimuli, we set the values between zero and one. We could naively scale the generated series to $[0,1]$, but this would invalidate our experiment. To understand why, assume without loss of generality that the averages of the base series are around 0.5 and that the generation procedure adds noise to larger y-values (for not mirrored series). Therefore, the generated series are between 0 and $\geq1$ (with the exact amount depending on the noise). Therefore, if we rescaled the data to $[0,1]$, we would push the average values of the series to lower y-values. 

Let us assume our participants respond randomly or always estimate the average at 0.5. In both cases, we get the same result. Since the true averages are overall lower, we would find that estimates are biased to be higher than the average. We believe that every experiment that investigates bias should test its analysis with random responses. With random responses, any observed human bias should disappear.

To overcome the issue, we scale \added{(multiply)} the averages of the low-noise data to have the same averages as the high-noise data\removed{ with a factor}.

\begin{equation}
    \text{scalingFactor} = \frac{\text{average}(\text{highNoiseData})}{\text{average}(\text{lowNoiseData)}}
\end{equation}

For mirrored series, we scale by $(1 - \text{average}(\text{highNoiseData})) / (1 - \text{average}(\text{lowNoiseData)})$. After this scaling, all stimuli with the same seed (and mirroring) have the same average. If we simulate an experiment where participants always estimate 0.5, we find no bias toward higher-variable areas.

\subsection{Experiment 2}

We use the same stimuli generation procedure as in Experiment 1 and the same seeds. In addition to the line encoding, we created graphs with points sampled at equal distances along the x-axis and sampled along the arc of the lines. This design resulted in three stimuli types (lines, Cartesian spaced points, and arc spaced points). Like in Experiment 1, we used two---albeit different---levels of variability in the graphs (no additional variability, and 0.4). We chose no additional variability for the first level to understand the generated series' baseline bias. We used the same level of variability for the high variability series to replicate Experiment 1. The series had 60 data points and were
scaled as in Experiment 1. \Cref{fig:exp_2_stimuli} shows example stimuli.


\section{Design, Procedure, and Participants}

\subsection{Design}

\textbf{Experiment 1:} We used a 2 (variability: .15 and .4) x 2 (variability upper vs. lower) within-subjects design. Average estimation error was collected as the dependent variable. This design resulted in a total of four trial types which were generated 12 times (48 total trials) to ensure test-retest reliability.

\vspace{0.4em}
\noindent\textbf{Experiment 2:} We used a 2 (variability: 0 and .4) x 2 (variability: upper vs. lower) x 3 (mark type: line graph, Cartesian spaced points, arc spaced points) mix-design. The between-subjects measure was mark type, and the within-subjects conditions were variability 0/.4 and variability upper/lower. Variability upper/lower was used as a manipulation check. Mean estimation error was collected as the dependent variable. Each participant completed the task with graphs that included variability 0/.4 and variability upper/lower in a randomized order. The total number of trials was the same as in Experiment 1. 

\subsection{Procedure}

In both experiments, participants completed this study online on their personal machines. After giving Institutional Review Board (IRB) approved consent to participate, individuals were given three types of instructions. The first set of instructions prompted participants to set their browser window to 100\% zoom. The second set of instructions pertained to the task, which was:

\vspace{0.5em}
\begin{addmargin}[1.5em]{1.5em}
``\textit{\textbf{Experiment Instructions}. 
Please read the following paragraphs carefully. You will be asked questions about the information in the paragraphs.}

\noindent
\textit{\textbf{Scenario}: Assume that you are a stock market investor. You are investing your own money in stocks, and you want to determine the average price of a stock over time in order to pick the best investment.}

\noindent
\textit{\textbf{Task}: In this experiment, you will be shown graphs of stock prices over a one-year period like the one below. Your task is to determine the average stock price for that year.}

\noindent
\textit{What is the average stock price? (Click and drag the line to indicate the average stock price)}

\noindent
\textit{\textbf{Response}: To indicate the average stock price, use your mouse to drag the line on the chart. Move the line to where you think the average stock price is for that year. You can readjust the line by clicking and dragging. Once you are happy with your judgment of the average stock price, click the next button.}''
\end{addmargin}

\vspace{2mm}

\noindent
The final set of instructions was an attention check, where participants were asked to fill in a blank with the word ``stock''. The sentence was, ``\textit{During this study, you will be asked to look at graphs of \_\_\_\_\_ prices.}'' Following the instructions, participants completed 48 estimation judgments in a randomized order. They indicated their judgments using a horizontal slider that was superimposed on the stimuli (shown in~\Cref{fig:prereg_exp}) to estimate the average data value in the graphs. The trials included text reminding the participants about the task. If participants failed to move the slider, they would be prompted to do so and restricted from progressing until they made their judgment. They received no feedback as to the accuracy of their judgments. 

 Following the main experiment, participants answered open ended questions about their strategy and what they thought the experiment was about. They also reported their gender and age. 


\subsection{Participants}

Based on the effect size calculated from pilot data, a power analysis was conducted using G*Power, to determine an adequate sample size, and preregistered. At an alpha of 0.05, power of 0.95, 4 predictors, and an effect size of adjusted r-square of 0.13, the minimum number of participants needed is 132, which we rounded to 140. For Experiment~1, participants were 142 people from Amazon's Mechanical Turk, with participation criteria set to workers in the US who were 18 years of age or older. Participants demographics were 98 male and 44 female, with an average age of 39 (\textit{SD} = 9). 

For Experiment~2, participants were 420 (140 per between-subjects group) people from Amazon's Mechanical Turk. Of those who chose to answer, 46\% identified as female, with an average age for the whole sample of 41 (\textit{SD} = 11). IRB approval for this research was obtained from (removed for anonymization) University's IRB. Participants were paid in accordance with (removed for anonymization) minimum wage. 

\section{Results}

To answer our primary analysis question, whether the perception of averages in lines is biased toward variability and whether we can manipulate and predict this bias, we will detail the results of the two experiments. In each experiment, we will begin with descriptive statistics about estimation error. We then show the results of statistical tests of our preregistered hypotheses. For all of the analyses, we did not remove any participants.

Following the preregistered analysis, we detail the thematic analysis of participants' strategies, including examining the variability-overweighting exhibited by participants who reported using the correct strategy. We also conducted a sensitivity analysis to determine if individuals who guessed the purpose of the study biased the results. We conclude the analysis with model comparisons that use the average along the arc to predict the observed biases.

\subsubsection{Accuracy Calculation}
We computed the error for each participant's estimates as the difference  between the estimated average and the true average.

\vspace{-.7em}
\begin{equation}
\text{Error} = \text{Estimated Average} - \text{Average}
\end{equation}

We also calculated whether a participant overestimates the average. We specify that they overestimated when the estimated average is higher than the true average of the time series data.

\vspace{-.7em}
\begin{equation}
\text{Overestimated} =
\begin{cases}
    \text{Overestimated},& \text{if } \text{Error} > 0\\
    \text{Underestimated}, & \text{otherwise}
\end{cases}
\end{equation}

To always have the high variability data at the higher y values, we also compute a normalized average as:

\vspace{-.7em}
\begin{equation}
\text{Normalized Average} = 
\begin{cases}
    -\text{Average} + 1,& \text{if } \text{variability upper vs. lower}\\
    \text{Average}, & \text{otherwise}
\end{cases}
\end{equation}

We similarly compute a normalized error where a positive error indicates an estimated average toward higher variability.



\subsection{Experiment 1}

\begin{figure}[ht]
  \centering
  \includegraphics[width=0.75\columnwidth]{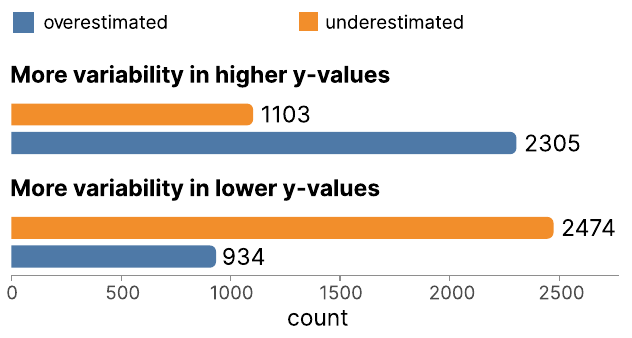}
 \caption{Counts of the over- and under-estimations in Experiment 1, broken down by the variability upper vs. lower condition.}
  \label{fig:counts1}
\end{figure}

\noindent
\textbf{Descriptive statistics}.
As a preliminary analysis of estimation error, we counted how many times participants over or underestimated the average. Of the 6816 responses, 3239 (48\%) were overestimated, and 3577 (52\%) were underestimated (see~\Cref{fig:counts1}). Participants generally underestimated averages.

Since our experimental data contains graphs with variability in the upper and lower y-values, we broke down these counts by this condition to determine whether the estimates are toward or away from the higher variability. As shown in \Cref{fig:counts1}, for graphs where the variability was in the higher y values, participants overestimated 2305 (68\%) trials, and they underestimated 1103 (32\%). In the condition where the variability was in the lower y values, 934 (27\%) were overestimated, and 2474 (73\%) were underestimated. In both conditions, the estimation error was consistent with the variability.

\begin{figure*}[t!]
  \centering   \includegraphics[width=0.98\linewidth]{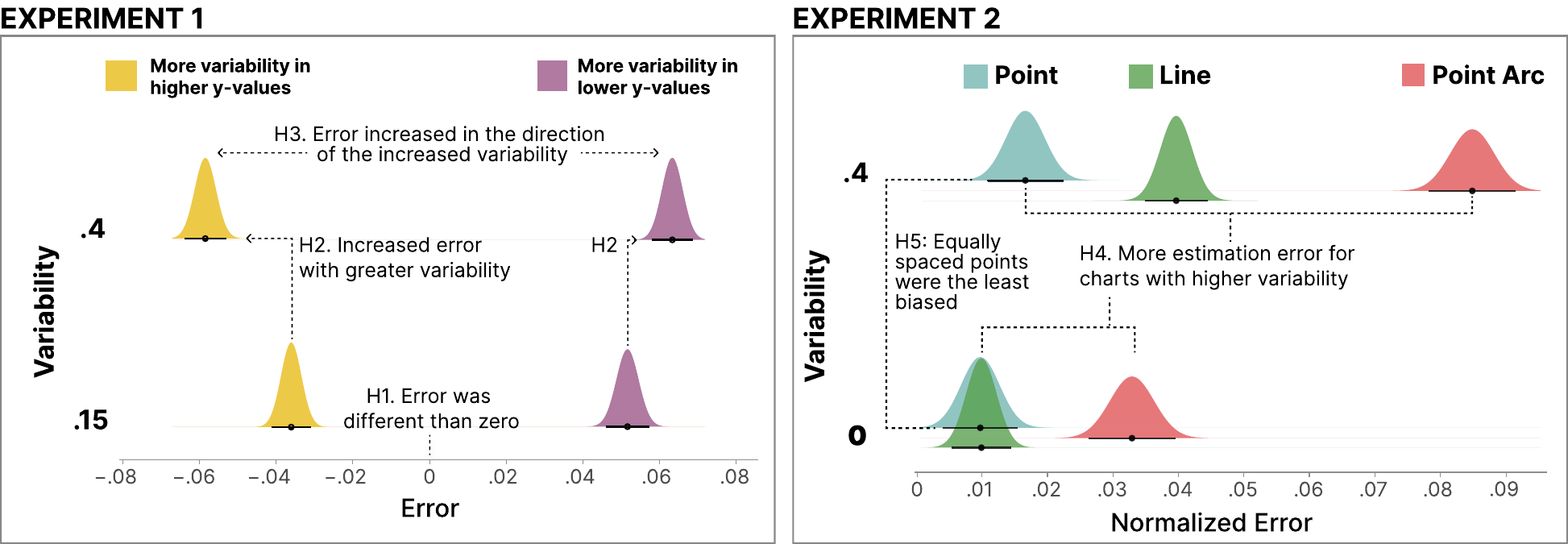}
 \caption{Experiments 1 and 2 results, showing the impact of variability, variability upper vs. lower, and mark type on estimation error. The left panel details the findings of Experiment 1 with annotations describing confirmed hypotheses 1-3. The right panel shows Experiment 2 with annotations describing confirmed hypotheses 4-5. \added{The black bars within the density plots show 95\% CIs with a mean dot.}}
  \label{fig:Exp1_plot}
\end{figure*}

\vspace{5mm}
\noindent
\textbf{Statistical tests of preregistered hypotheses.} 
To determine if the findings from the descriptive statistics are robust, we conducted a statistical analysis of our preregistered hypotheses.
We preregistered three hypotheses on the Open Science Framework for the first experiment\footnote{In the original pre-registration, we used the term \textit{noise} rather than \textit{variability}. We updated the term  here to be consistent.}. 

\begin{enumerate}
 \item[H1] \textit{``Estimation error will be significantly different than zero.''}
 \item[H2] \textit{``There will be significantly more estimation error for trials with higher variability compared to lower variability.''} 
 \item[H3] \textit{``Estimation error will be observed in the direction of the increased variability (\ie positive errors will be observed when the area of highest variability is above the average y-value and negative errors will occur when the highest variability is lower on the y-axis than the average.)''} 
\end{enumerate}

To test these hypotheses, we conducted the preregistered analysis, in which a linear regression model was fit to the data using the R function lmer~\cite{r} with restricted maximum likelihood estimation procedures~\cite{raudenbush2002hierarchical}. \added{Note that we used multi-level linear regression models to account for correlations between participants' responses instead of the more simplistic pre-registered linear regression models. } Linear regression assumptions were tested and met. 
The model included \textit{variability size} (.15 vs. .4), \textit{variability position} (upper vs. lower), their interaction, and random intercepts for each participant to predict errors in participants' average estimations. The referents were .15, and variability in the upper y-values. \added{The resultant model in R notation was: \(Error \sim variabilitySize * variabilityPosition + (1 | Id)\).}


\vspace{.5em}
\noindent
\textbf{Test of H1: \textit{estimation error will be significantly different than zero}.} The results revealed a significant intercept of the model ($\textit{b} = -0.036$, $\textit{t}(6,811) = -6.6$, $\textit{p} < .001$, 95\% CI $[-.047, -.025]$), providing evidence that the absolute estimation error (3.6\%) for the referent conditions was meaningfully different than zero (supporting H1). This effect can be seen in~\Cref{fig:Exp1_plot} (left panel), which displays estimation errors for each condition, with none of the conditions overlapping zero.

\vspace{.5em}
\noindent
\textbf{Test of H2: \textit{significantly more estimation error for trials with higher variability.}} The results also revealed a significant main effect of variability ($\textit{b} = -.022$, $\textit{t}(6,811) = -6.5$, $\textit{p} < .001$, 95\% CI $[-.029, -.016]$). This effect can be seen in \Cref{fig:Exp1_plot}, where there is a meaningful separation
between the two variability types for the variability upper vs. lower
conditions (denoted with H2). 
This finding supports H2, suggesting significantly more estimation error for trials with higher variability than lower variability. 

\vspace{.5em}
\noindent
\textbf{Test of H3: \textit{estimation error will occur in the direction of the increased variability.}}
There was also a significant interaction between \textit{variability .15 vs .4} and \textit{variability upper vs. lower} ($\textit{b} = .034$, $\textit{t}(6,811) = 7$, $\textit{p} < .001$, 95\% CI $[.025, .044]$). To unpack the interaction, we ran the same model as above but with the variability in the lower y-value graphs as the referent. This model yielded a significant effect of variability but in the opposite direction ($\textit{b} = \textbf{.012}$, $\textit{t}(6,811) = 3.39$, $\textit{p} = .001$, 95\% CI $[.005, .018]$) compared to the prior model ($\textit{b} = \textbf{-.022}$). As seen in \Cref{fig:Exp1_plot}, errors occurred in the direction of the increased variability, supporting H3. We found positive errors when the area of highest variability was above the average y-value and negative errors when the highest variability was lower on the y-axis than the average. 






\subsection{Experiment 2}
To examine one possible source of the variability-overweighting, in Experiment 2, our goal was to identify the contribution of the line encoding. We predicted that there would be an interaction between variability and the mark type, such that the effect of variability will be smaller for graphs with points spaced along the x-axis than graphs with points spaced along the arc and line graphs.

\begin{figure}[ht]
  \centering
  \includegraphics[width=0.7\columnwidth]{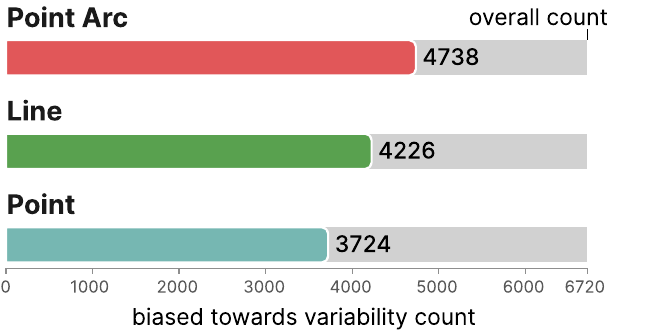}
 \caption{Counts of the number of estimates that were biased toward variability in Experiment 2 for each mark type.}
  \label{fig:counts2}
\end{figure}

\vspace{3mm}
\noindent
\textbf{Descriptive statistics}.
Using the same methods as in Experiment 1, we counted how many times participants were biased toward variability (see \Cref{fig:counts2}).
Of the 6720 responses, 3724 (55\%) for Point, 4226 (63\%) for Line, and 4738 (71\%) for Point Arc \added{were biased toward variability}. 

\vspace{3mm}
\noindent
\textbf{Statistical tests of preregistered hypotheses.} 
To test the reliability of the descriptive statistics, we preregistered two hypotheses on the Open Science Framework for the second experiment.

\begin{enumerate}
 \item[H4]  \textit{``There will be significantly more estimation error for trials with higher variability than no additional variability.''}  
\item[H5]\ \textit{``The least variability-overweighting will occur in graphs with points that are equally spaced along the x-axis.''}
\end{enumerate}

We used a multilevel model to fit the data using the \textit{lmer} package~\cite{lmer} in R, which is appropriate for mixed designs with between- and within-subjects variables. The model used variability (0 and .4) to predict normalized estimation error (testing H4). We calculated the normalized estimation error for each condition using the absolute error (the error is computed in the same way as in Experiment 1) when the graph was vertically mirrored.  We used \textit{normalized error} rather than \textit{error} and removed the \textit{variability upper vs. lower} term to reduce the complexity of the model, which was preregistered. To evaluate H5, we also included an interaction term between mark type and variability and the necessary lower-order terms. Finally, we included random intercepts for each participant. The resultant model in R notation was: \(NormalizedEstimationError \sim markType * variability + (1 | Id)\). The referents of the model were the line mark and zero variability. 


\begin{figure}[ht]
  \centering
  \includegraphics[width=0.8\columnwidth]{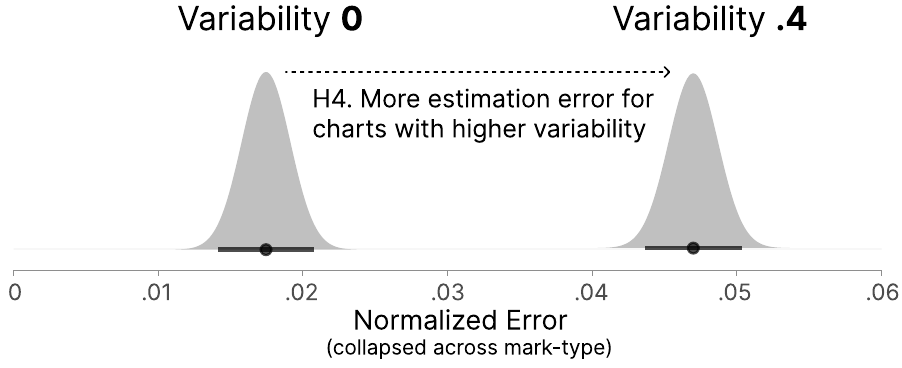}
 \caption{The meaningful main effect of variability in Experiment 2, averaged over the mark types, including annotations describing confirmed H4. \added{The black bars within the density plots show 95\% CIs with a mean dot.}}
  \label{fig:Exp2main}
\end{figure}  

\vspace{.5em}
\noindent
\textbf{Test of H4: \textit{significantly more estimation error for trials with higher variability.}}
Replicating Experiment 1, the model results revealed a main effect of variability ($\textit{b} = 0.03$, $\textit{t}(20,153) = 10.03$, $\textit{p} < .001$, 95\% CI $[.024, .036]$), indicating that graphs with more variability had greater estimation error. \Cref{fig:Exp2main} shows that, when collapsed across the mark types, estimation error increased by .03 from graphs with no additional variability to .4 variability (confirming H4). The meaningful increase in error with the more variable graphs can also be seen in \Cref{fig:Exp1_plot} (right panel), which shows the impact of variability on each mark type. 


\vspace{.5em}
\noindent
\textbf{Test of H5: \textit{least variability-overweighting in graphs with equally spaced points.}} As shown in \Cref{fig:Exp1_plot} (right panel), Point (Cartesian spaced) had the smallest change in normalized error from charts with low to high variability (0 to .4). 
Our results revealed the change from low to high variability
 was meaningfully larger for Line vs. Point ($\textit{b} = -.023$, $\textit{t}(20,153) = -5.45$, $\textit{p} < .001$, 95\% CI $[-.031, -.015]$). The change in normalized error from low to high variability was also meaningfully larger for Point Arc vs. Point ($\textit{b} = .05$, $\textit{t}(20,153) = 10.74$, $\textit{p} < .001$, 95\% CI $[-.031, -.015]$). Point Arc showed the largest increase in the normalized error of 5\%, followed by Line (3\%), and then Point (.69\%).

We conducted a follow-up regression analysis to determine if the small increase in normalized estimation error was meaningful for Point. This analysis revealed a meaningful but small bias for Point ($\textit{b} = .007$, $\textit{t}(6,718) = 2.24$, $\textit{p} = .025$). In sum, these results support H5, indicating that the points equally spaced along the \added{x-axis} had the least bias (.69\%), with the line (3\%) and points along the arc showing greater bias (5\%). 


\definecolor{WildSand}{rgb}{0.96,0.96,0.96}
\definecolor{Gray}{rgb}{0.501,0.501,0.501}
\begin{table*}
\small
\centering
\caption{The six most reported strategies for each experiment and the proportion of participants who correctly guessed the purpose of the study. The last column shows the inter-rater reliability score (IRR), which indicates the level of agreement between raters.  IRR scores over .61 indicate substantial agreement between raters~\cite{mchugh2012interrater}.}
\label{tab:strategies}
\vspace{-.5em}
\begin{tblr}{
  cells = {c},
  column{2} = {WildSand},
  column{3} = {WildSand},
  column{4} = {WildSand},
  column{5} = {WildSand},
  column{6} = {WildSand},
  column{7} = {WildSand},
  cell{1}{2} = {c=6}{},
  hlines,
  hline{1,9} = {-}{0.08em},
  hline{4-8} = {-}{Gray},
}
& \textbf{Reported Strategy} & & & & & & { \textbf{Guessed purpose} \\ \textbf{of study} } \\
Exp & {mental \\ \textbf{averaging} } & {focusing on \\ \textbf{extrema}} & {incorporated \\ \textbf{variability}} & {equal number of \\ \textbf{points} or \textbf{line} \\ below and above} & {equal \textbf{area} \\ below and above} & {\textbf{beginning} and \\ \textbf{end} points } & {variability-\\overweighting} \\
{Exp 1\\Line}      & 56.74\%                    & 21.28\%                    & 21.99\%                  & 5.67\%                                                         & 3.55\%                                      & 4.96\%                              & 2.13\%                                \\
{Exp 2\\Line}      & 52.42\%                    & 23.39\%                    & 17.74\%                  & 8.87\%                                                         & 5.65\%                                      & 2.42\%                              & 4.03\%                                \\
{Exp 2\\Points}    & 38.13\%                    & 9.35\%                     & 10.07\%                  & 24.46\%                                                        & 1.44\%                                      & 2.16\%                              & .72\%                                 \\
{Exp 2\\Point Arc} & 32.85\%                    & 18.25\%                    & 12.41\%                  & 7.30\%                                                         & 5.11\%                                      & 1.46\%                              & 1.46\%                                \\
\textbf{Average}          & \textbf{44.92\%}                    & \textbf{17.93\%}                    & \textbf{15.53\%}                  & \textbf{11.65\% }                                                       & \textbf{3.88\%}                                      & \textbf{2.77\% }                             & \textbf{2.03\% }  
\\ 
\textit{\textcolor[rgb]{0.498,0.498,0.502}{IRR}} & \textcolor[rgb]{0.498,0.498,0.502}{.83}                              & \textcolor[rgb]{0.498,0.498,0.502}{.88}                            & \textcolor[rgb]{0.498,0.498,0.502}{.70}                                     & \textcolor[rgb]{0.498,0.498,0.502}{.87}                                                                            & \textcolor[rgb]{0.498,0.498,0.502}{.89}                                       & \textcolor[rgb]{0.498,0.498,0.502}{.83}                                              &  

\end{tblr}
\vspace{-.5em}
\end{table*}

\subsection{Open Responses for Experiments 1 and 2}\label{section:open}
After completing the estimation judgments, participants answered an open-ended question about their strategies in the task. The question was, ``\textit{We are very interested in how you made your decisions about the average stock price. Please list all the things you considered when making your judgments}.” Two raters read the responses and coded them based on the six most common strategies to analyze these data. The following sections report the six most frequent strategies and include example responses. In this analysis, we identify that a small proportion of the participants reported using the correct strategy. We conducted a follow-up analysis to determine if those who were consciously aware of the correct strategy displayed less variability overweighting.

Participants also reported their beliefs concerning the purpose of the experiment to determine if any participants intentionally biased their judgments. The question text included,``\textit{What do you think the experiment was about?}'' In the second part of this section, we report the proportion of participants who were aware of the purpose of the study. Then we conducted a sensitivity analysis to determine if the people who guessed the purpose of the study meaningfully impacted the findings. 

\begin{figure}[ht]
  \centering
  \includegraphics[width=1\columnwidth]{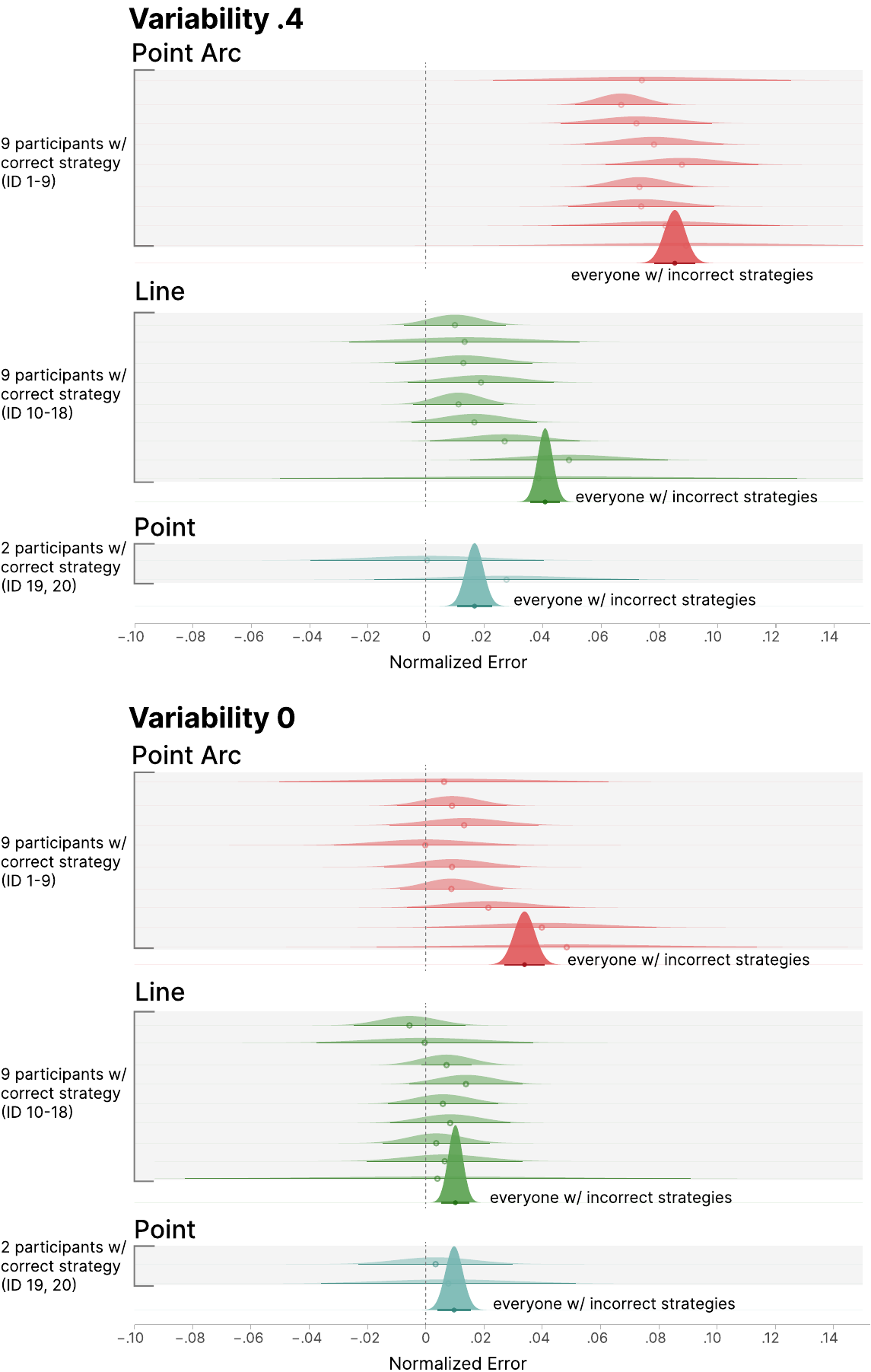}
 \caption{Strategy analysis for Experiment 2, where individual participants with the correct strategy (gray background) are shown compared to the other participants (foreground density plots). These data are broken down by mark type and variability. \added{The horizontal bars within the density plots show 95\% CIs with a mean dot. The dashed line denotes zero normalized error. }}
  \label{fig:Exp2strg}
\end{figure}

\subsubsection{Reported strategies}\label{sec:strategies}
We identified six main strategies that participants used to estimate the average of the stock data. Using the strategy codes,  we computed inter-rater reliability scores (\textit{IRR}, Cohen's Kappa)~\cite{irr2019} for the codes to determine the level of agreement between the raters (shown in the bottom row of \Cref{tab:strategies}). The average inter-rater reliability  for the six questions was .83 and ranged from .70 to   
.89. This range of inter-rater reliability scores indicates a substantial level of agreement between the two raters~\cite{mchugh2012interrater}. 
The codes were not mutually exclusive, and many times participants indicated that they used several strategies, in which case they received multiple codes. The proportion of strategies reported in \Cref{tab:strategies} is for the codes the two raters agreed on.

\textbf{Mental averaging.} The most commonly reported strategy was mentally computing the average using visual perception. For example, a participant wrote, ``\textit{I marked the point on the graphs where it seemed like the generalized average would fall if the points on the graph were boiled down into numbers and you wanted to find the average of those numbers.}'' Another participant described, ``\textit{I tried to get a visual sense of where the average would fall. I looked for a good mid point of the overall graph.}'' \Cref{tab:strategies} shows that this was the most
commonly reported strategy for each mark type. 

\textbf{Focusing on extrema.} The second most common strategy was to focus on the max and min points and select a location between those extrema. For example, a participant wrote, ``\textit{I looked at the highest and lowest point and went with the middle.}'' Another example includes, ``\textit{I looked at the lowest mark and the highest mark and then the middle of that, but looked to see if there were upper or lower trends and adjusted accordingly to that}...'' Focusing on the extrema is not the most effective strategy. It is surprising to see that, on average, roughly 17\% of the participants indicated that they incorporated the high and low points into their average estimations. 

\textbf{Incorporating variability.} Roughly 15\% of participants reported incorporating variability into their judgments. However, they incorporated the variability in different ways. For example, one type of strategy included incorporating the areas with both high and low variability. For example, a participant wrote, ``\textit{I tried to find out the relatively stable parts of the graph, these were useful when they extended over a long period of time. I also considered the effect of the crests and troughs and the depth of these extreme occurrences. Using these as a metric, I tried to estimate the average.}'' 

In contrast, another group of participants focused more on areas with low variability. For example, ``\textit{Most of the time the stock price comes to certain point, and jumps again or fall back, I consider the price where it is often stable for more time.}'' or ``\textit{It was easier to make the average when the stock prices were not changing much and the graph was more even. When the prices were more ``jumping'', I tried to find the phase where these trends stayed the longest and put my average around it.}'' Participants who reported this type of strategy seemed averse to variability or uncertainty, which is a well-known bias in psychology~\cite{riskaverstion, fox1995ambiguity}. 

\textbf{Equal number of points or line below and above.} 
A strategy we did not anticipate was ensuring an equal number of points or line lengths above and below the judgment indicator. To indicate their responses, participants drug a line on the graph. By allowing participants to place a line directly on the graphs, participants could then easily count the number of points above and below the line. One participant simply wrote, ``\textit{I tried to have the number of points above and below the line be approximately equal}.” Unsurprisingly, this strategy was most common for participants who viewed graphs with points equally spaced along the x-axis (24\%). Although less common, some participants who viewed the line encoding also used this strategy (5-8\%). A participant explains, ``\textit{I tried to get half of the trend line above and half of the trend line below the average line and where I placed it}.” 

\textbf{Beginning and endpoints.} 
Another suboptimal strategy was to focus on the beginning and ending values of the time series. While a small proportion of participants used this strategy (roughly 3\%), it is noteworthy because it reflects a misconception. For example, one participant wrote, ``\textit{I mainly looked at the stock at the beginning and end of the year. Afterwards, I tried to make an educated guess on what the average stock price would be}.'' Another person describes also being confused about the impact of data at the end of the time series. They wrote, ``\textit{Depending on how the end of the chart looks, I draw a different strategy. If the chart is rallying, I believe the average price is at the low before this rally. If the chart is going down, I place it at the lowest low there was throughout the chart}.''

\textbf{Equal area (correct strategy).} 
The correct strategy was to select a location with equal \textit{area} above and below the estimated average. Only a small number of people reported using this strategy (roughly 4\%). An example is, ``\textit{I just tried to make the volume of the areas above and below the line approximately equal. That was my only strategy. Think I learnt it in a maths or stats course.}'' 

As the equal-area strategy is the correct approach, we wanted to determine if participants who used it showed less bias in their judgments. To compare performance between those who used the equal-area strategy to those who did not, we computed the bias for the two groups for Experiment 2. \Cref{fig:Exp2strg} shows the nine participants with the correct strategy in the Point Arc and Line groups and the two in the Point group compared to a distribution representing all the other strategies. Note that there is one distribution for all people with the incorrect strategies compared to individual distributions for those with the correct ones. We did this to clarify that a small number of people had the correct strategy and meaningful variation exists between them. 

Taking the individual distributions from those with the correct strategy as a whole compared to the distributions for those with the incorrect strategy, we found that people with the correct strategy showed 12.7\% less bias (.028 normalized error) than those with the incorrect strategy (.032 normalized error). The disparity between those with the correct strategy (.021 normalized error) and without (.041 normalized error) was most pronounced for the Line encoding with .4 variability (change of .019 or 46\% reduction).  We opted not to do a statistical analysis on these groups as they were highly unbalanced (20 participants vs. 398) and were not equally distributed across the groups. However, visual analysis reveals a general tendency where using the correct strategy leads to less bias.  

\subsubsection{Knowledge about the experiment purpose} 
Several people in each experimental condition made guesses somewhat close to the actual experiment goals in response to the question, ``What do you think the experiment was about?'' For example, one person wrote, ``\textit{How people picture averages differently when there are smooth transitions versus spikes in the graph.}'' and another person wrote ``\textit{I think it was about how accurate people can estimate the average of a line and if different line conditions affect the accuracy, such as jagged line vs. smooth line...}''

\begin{figure}[ht]
  \centering
  \includegraphics[width=\columnwidth]{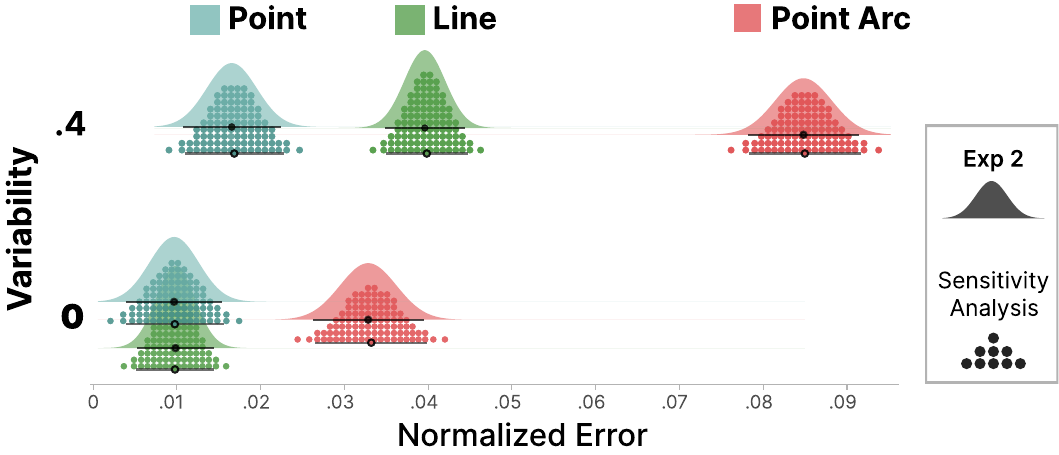}
 \caption{Sensitivity analysis for Experiment 2, in which the original data from Experiment 2 is displayed with density plots, and the data that excludes people who guessed the purpose of the experiment is shown with quantile dotplots. \added{The black bars within the density plots show 95\% CIs with a mean dot.}}
  \label{fig:Exp2guessed}
\end{figure}

In Experiment 1, three people guessed the purpose of the study, and ten correctly guessed in Experiment 2. We conducted a sensitivity analysis to determine if those participants biased the findings. In this analysis, we removed the participants that relatively accurately guessed the manipulations of the study and reran the preregistered analysis for Experiment 2. Across all the findings, there was no meaningful impact of removing the participants who guessed correctly. To illustrate these effects, in \Cref{fig:Exp2guessed}, we show the original data from Experiment 2 with density plots. Overlayed on the density plots are quantile dot plots that show the data after removing the participants who guess the manipulations in the study. As seen in \Cref{fig:Exp2guessed}, where all the distributions for each condition overlap, removing the participants did not meaningfully impact the results. 

\subsection{Predictive Model} 
\label{sec:predictive-md}

\added{Chart authors should consider different designs if a particular line chart is prone to bias.
To help chart authors know whether a chart may be biased without running their own perceptual experiments, we sought to build a model that predicts participants' responses.}
We aim to predict the bias and average estimate only based on properties of the data we can observe in a given line chart \removed{(such as the true average and arc average)} rather than based on the parameters of the data generation method since the latter is typically not known. We also chose to only use a simple model with few features (rather than, \eg the whole time series as input) since we are interested in the model's generalizability.

\added{We hypothesize that such a model is possible.}
If the salience of longer line segments drives the estimates of averages, we may be able to predict the estimates of averages using the true average and the average of the values along the arc---\emph{arc average} for short.


To understand whether the arc average is a meaningful predictor, we computed the Pearson correlation between the average error of the average estimate for each stimulus to the error of the arc average estimate. For Experiment 1, the correlation is 0.85 ($\textit{p} < .001$), and for Experiment 2, the correlation is 0.64 ($\textit{p} < .001$). This suggests that the arc average is meaningful to predict the variability overweighting bias.

\removed{In light of the prior evidence, that the arc average helps predict the bias, we sought to model the estimated average.} To predict the estimated average we created \removed{a sequence of} linear regression models that first used the average of the data points to predict participants' responses and then a second model that included the arc average. The goal of the second model was to evaluate if the arc average accounted for meaningfully more variability in participants' responses than the average of the data set alone. We then statistically compared the two models to determine if the arc average model was a significantly better fit, using the data from Experiments 1 and 2.

For Experiment 1, we fit a linear regression model using the average of the data point to predict participants' estimates.
\removed{The results revealed that} The average of the data points meaningfully predicted participants' responses ($\textit{b} = .53$, $\textit{t}(6814) = 51.80$, $\textit{p} < .001$) with a model adjusted r-squared of .28.
For the second model that included arc average, both the average of the data points ($\textit{b} = .45$, $\textit{t}(6813) = 39.42$, $\textit{p} < .001$) and arc average ($\textit{b} = .25$, $\textit{t}(6813) = 15.70$, $\textit{p} < .001$) meaningfully accounted for variance in participants judgments.
The second model had an adjusted r-squared of .31. This result suggests that after accounting for the meaningful impact of the average of the data points, for every one unit change in arc average, participants' judgments were biased by .25. 
We then compared the two models using an ANOVA\removed{ equation}. This comparison revealed that the second model, which included arc average, had a significantly better fit than the first model ($\textit{F}(2, 6813) = 246.58$, $\textit{p} < .001$).

We also completed the same sequence of model comparisons for the data in Experiment 2 using only the data for the line stimuli.
For the first model, the impact of the true average was $\textit{b} = .69$, $\textit{t}(6718) = 76.07$, $\textit{p} < .001$, with an adjusted r-squared of .46.
For the second model, the effect of the true average ($\textit{b} = .39$, $\textit{t}(6717) = 11.62$, $\textit{p} < .001$) was larger than the impact of the arc average ($\textit{b} = .34$, $\textit{t}(6717) = 9.18$, $\textit{p} < .001$), with an adjusted r-squared of .47. This result suggests that after accounting for the meaningful impact of the true average, for every one unit change in arc average, participants' judgments were biased by .34 (compared to .25 from Experiment 1).
When comparing the two models, we found that the model that included the arc average had a meaningfully better fit than the one that did not ($\textit{F}(2, 6717) = 84.30$, $\textit{p} < .001$).

\begin{figure}[ht]
  \centering
  \includegraphics[width=\columnwidth]{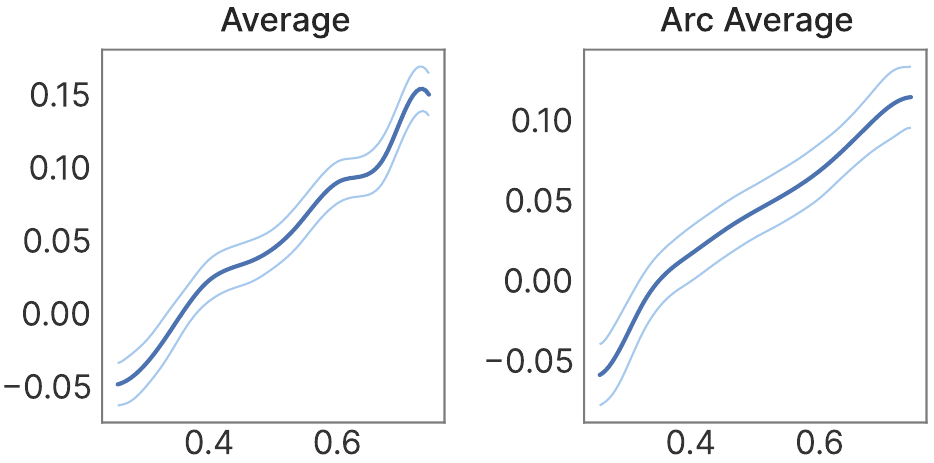}
  \caption{The two response functions of a generalized additive model (GAM) trained on the line data from Experiments 1 and 2. The response functions resemble linear functions, making linear models appropriate for these data.}
  \label{fig:gam}
\end{figure}

We then created a linear model from the data for both experiments with three parameters: intercept ($\textit{b} = .14$, $\textit{t}(13533) = 32.58$ $\textit{p} < .001$), average ($\textit{b} = .40$, $\textit{t}(13533) = 34.36$, $\textit{p} < .001$), and arc average ($\textit{b} = .31$, $\textit{t}(13533) = 21.87$ $\textit{p} < .001$), and an adjusted r-squared of .39. We selected a linear model because we found that a more sophisticated GAM~\cite{hastie2017generalized} used nearly linear feature functions~(\Cref{fig:gam}).

\begin{figure}[ht]
  \centering
  \includegraphics[width=0.7\columnwidth]{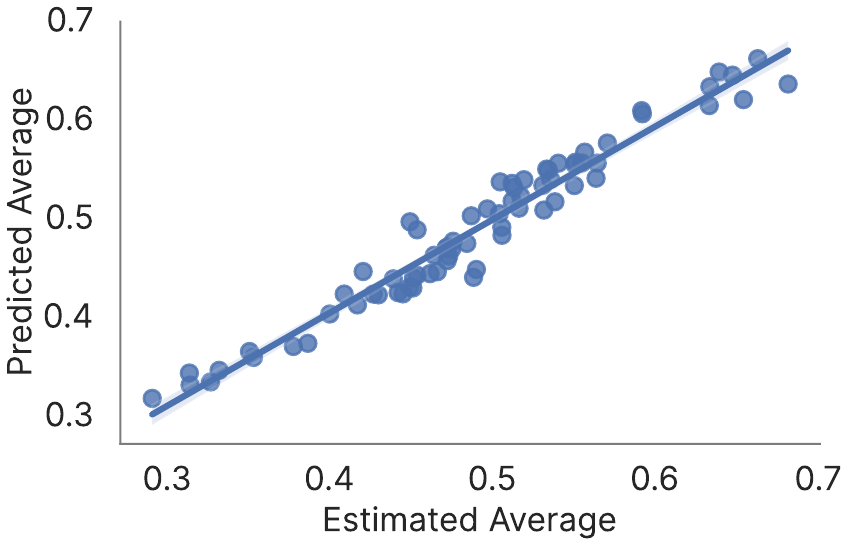}
  \vspace{-.5em}
  \caption{\removed{Comparison of the predicted average and the estimated average for all stimuli.}\added{Predicted and estimated averages for all stimuli.}}
  \label{fig:modelfit}
\end{figure}

\removed{We evaluated whether the model accurately predicts the average responses for all our stimuli.}
For most stimuli (90\%), the model predicted the correct direction of the bias. The predicted estimated average fits well with the estimated averages~(\Cref{fig:modelfit}). The model's mean absolute error is .014, which means that with values in the range of [0,1], the prediction is only off by 1.4\%. The RMSE is .019, also indicating a good fit.

\section{Discussion}

The results of Experiment 1 (\Cref{fig:Exp1_plot}, left) support our hypothesis that estimation error is significantly biased toward the direction of larger variability in data.
This bias could be caused by more variability leading to steeper line segments that use more ink and are more visually salient. Prior work has also found that areas of higher salience in scatter plots can bias average estimation~\cite{hong2021weighted}.
To test this theory, we conducted Experiment 2, using a dot plot instead of a line chart to encode the series data.
In the dot plot, the amount of salience is proportional to the amount of data in each x-interval and independent of the steepness of the line segment.

Experiment 2 (\Cref{fig:Exp1_plot}, right) replicates the findings from Experiment 1.
The experiment additionally supports our hypothesis that we can reduce the bias by encoding the series data as a dot plot instead of a line chart.
To simulate the higher salience of steep line segments, we also tested a design that spaces points along the arc of a line.
We found that the line bias was significantly higher than the bias of the dot plot but lower than the average estimation of the points along the arc.
These results support our theory that the bias is toward more visually salient areas of the chart but cannot yet explain the full extent of the bias.

We generated the stimuli for our experiments using different levels of variability. Since these parameters are typically unknown, it would be impossible to model the bias in real-world applications.
However, if we assume that the bias is caused by the salience of steep line segments, we can compute the direction of the bias directly from the average of the points along the arc of the line.
We can also estimate the magnitude of the bias as a function of the average of the series and the average of the points along the arc using a simple regression model (\Cref{sec:predictive-md}).
Future work could refine this model using more features\removed{ sophisticated modeling techniques}.

Our experiments show that average estimates are biased toward higher variability. We believe that we could similarly bias trend estimation in line graphs. For example, a line graph could have more variability for smaller values in the first half and higher variability for larger values in the second half. Since we found that the estimates of averages for the first half are lower than the true average and that estimates for the second half are higher, we can expect that a person also perceives a more extreme increase (stronger trend) than there is. Our experiment only investigated average estimation in isolated charts. As such, future work must confirm that this bias exists in combined charts. Suppose trend estimation could be biased by variability. In that case, malicious people who can affect the variability of series could influence decisions other people make based on trends in data, such as in stock trading.

The results of our experiments have implications for the design of charts in applications where people estimate the average or trend of a series.
Designers should consider whether they can replace a line chart with a dot plot to reduce the bias.
\added{However, the dot plot design makes the data order and the delta between consecutive points less clear.}
There may also be other ways to reduce the bias, such as using a different type of line chart that de-emphasizes steep line segments using thinner line segments or lines with lower opacity.

We asked participants about their strategies for estimating the average and found that people used a variety of strategies, the majority of which were incorrect. The high proportion of misconceptions observed in participants' strategies is concerning. 
We also found that those who used the correct strategy of aiming for an equal area between the average line and the data line \added{seemed to be} less biased (\Cref{sec:strategies}).
While we have too few people to draw firm conclusions, this insight suggests that people \added{may be able to} learn to reduce the bias by using the correct strategy. \added{When we initially developed this work, we hypothesized that the biases would be driven by visual salience, a bottom-up attentional process. While such unconscious processes are certainly part of the cause, these data provide some indication that strategies may play a role. One limitation of this work is that we cannot disambiguate the effects of visual salience and strategies. The interconnection between the two is consistent with theories in visual attention that suggest strategies and bottom-up processes are intrinsically interconnected, forming a feedback loop~\cite{salience, padillamodel}. It is also possible that both the mark type and response method bias participant strategies, which could have impacted attention and responses.} Despite these limitations, our findings point to the possibility that visual literacy training \added{might benefit from} teaching people to use the correct strategy.

We carefully designed the stimuli of the experiment such that simulated random responses did not show the bias we expected in real responses (\Cref{sec:scaled-stimuli}).
We only found this subtle issue after some initial data generation and pilots and would therefore encourage everyone who runs experiments that test human biases to test their experiments with random data.

\section{Conclusion and Future work}

This paper shows that average estimates are biased toward areas of higher salience, caused by increased variability in the visualized data. Since this bias can affect the conclusions drawn from data, visualization designers who create line graphs must be aware of it. This bias is not only significant but also practically relevant. The amount of variability in a line graph may be due to irrelevant (to the conclusions) factors, such as inconsistencies in the data collection, such as varying sensor noise. In the worst case, a malicious actor could introduce small amounts of noise to mask larger changes or nudge analysts to \added{see} \removed{imagine} larger changes.

By quantifying the bias, we can consider showing viewers warnings when we expect the bias to affect the conclusions drawn from a graph or consider alternative visual encodings that do not have the bias shown in this paper. For example, we showed how points instead of lines reduce the bias. Another idea could be to reduce the salience of steep lines by varying the opacity or line width. Alternatively, designers could consider annotating graphs with averages or other visual encodings when average estimates are needed. However, designers need to consider potential biases that additional visual encodings could introduce.

We discussed that biased average estimates could also lead to biased trend estimates. Future work should investigate how manipulations of time series data visualized as line graphs affect trend estimates. Participants in our study represent a general population of people with some but not expert-level visualization literacy. If trend estimates can be affected, we should also investigate whether experts such as scientists, doctors who look at vitals, and stock traders are as affected as the general population. An avenue for investigating this effect could be to analyze historical data such as stocks and see whether increased variability affected traders' investments.

\acknowledgments{
This work was supported in part by grants from the NSF (\#2238175 and \#1901485).
}

\bibliographystyle{abbrv-doi-hyperref}

\bibliography{paper}
\end{document}